\documentclass[sigconf, nonacm]{acmart}

\newcommand\vldbdoi{XX.XX/XXX.XX}

 \newcommand\vldbavailabilityurl{URL_TO_YOUR_ARTIFACTS} 
\newcommand\vldbpagestyle{plain}

\usepackage{comment}
\usepackage{algorithm}
\usepackage{algpseudocode}
\usepackage{color}
\usepackage{colortbl}
\usepackage{multirow}
\usepackage{graphicx} 
\usepackage{subcaption} 

\usepackage{amsmath}
\usepackage{enumitem}
\usepackage{comment}
\usepackage{xcolor}
\usepackage{booktabs} 
\usepackage{url} 
\usepackage{stfloats}

\begin{document}


\title{Implementation and Privacy Guarantees for Scalable Keyword
Search on SOLID-based Decentralized Data with Granular
Visibility Constraints}

%
\author{Mohamed Ragab}
\orcid{}
\affiliation{%
  \institution{School of Electronics and Computer Science, University of Southampton}
  \city{Southampton}
  \country{United Kingdom}
}
\email{ragab.mohamed@soton.ac.uk}

\author{Faria Ferooz}
\affiliation{%
  \institution{School of Electronics and Computer Science, University of Southampton}
  \city{Southampton}
  \country{United Kingdom}
}
\email{f.ferooz@soton.ac.uk}

\author{Mohammad Bahrani}
\affiliation{%
  \institution{School of Electronics and Computer Science, University of Southampton}
  \city{Southampton}
  \country{United Kingdom}
}
\email{m.bahrani@soton.ac.uk}

\author{Helen Oliver}
\orcid{0000-0003-1467-8165}
\affiliation{%
  \institution{School of Computing and Mathematical Sciences, Birkbeck, University of London}
  \city{London}
  \country{United Kingdom}
}
\email{h.oliver@bbk.ac.uk}

\author{Thanassis Tiropanis}
\affiliation{%
  \institution{School of Electronics and Computer Science, University of Southampton}
  \city{Southampton}
  \country{United Kingdom}
}
\email{t.tiropanis@soton.ac.uk}

\author{Alexandra Poulovassilis}
\affiliation{%
  \institution{School of Computing and Mathematical Sciences, Birkbeck, University of London}
  \city{London}
  \country{United Kingdom}
}
\email{a.poulovassilis@bbk.ac.uk}

\author{Adriane Chapman}
\affiliation{%
  \institution{School of Electronics and Computer Science, University of Southampton}
  \city{Southampton}
  \country{United Kingdom}
}
\email{adriane.chapman@soton.ac.uk}

\author{George Roussos}
\affiliation{%
  \institution{School of Computing and Mathematical Sciences, Birkbeck, University of London}
  \city{London}
  \country{United Kingdom}
}
\email{g.roussos@bbk.ac.uk}

\begin{abstract}
In decentralized personal data ecosystems grounded in architectures such as \texttt{Solid}~\footnote{https://solidproject.org/}
, users retain sovereignty over their data via personal online data stores (pods), hosted on Solid-compliant server infrastructures. In such environments, data remains under the control of pod owners, which complicates search due to distribution across numerous pods and user-specific access constraints.
ESPRESSO is a decentralized framework for scalable keyword-based search across distributed Solid pods under user-defined visibility policies. It addresses key challenges of decentralized search by constructing WebID-scoped indexes within pods and employing privacy-aware metadata to enable efficient source selection and ranking across servers.
This paper further introduces a formal threat model for ESPRESSO, analysing the security and privacy risks associated with the generation, aggregation, and use of indexes and metadata. These risks include unintended metadata leakage and the potential for adversaries to infer sensitive information about data that reside within personal data stores. The analysis identifies key design principles that limit metadata exposure while mitigating unauthorized inference.
The proposed threat model provides a foundation for evaluating privacy-preserving decentralized search and informs the design of systems with stronger privacy guarantees.

\end{abstract}
\maketitle

\pagestyle{\vldbpagestyle}
\begingroup
\renewcommand\thefootnote{}\footnote{\noindent
This work is licensed under the Creative Commons BY-NC-ND 4.0 International License. Visit \url{https://creativecommons.org/licenses/by-nc-nd/4.0/} to view a copy of this license. Copyright is held by the owner/author(s).
%
}\addtocounter{footnote}{-1}\endgroup

\ifdefempty{\vldbavailabilityurl}{}{
\vspace{.3cm}
}

\section{Introduction}
In decentralized personal data ecosystems~\cite{abiteboul2015managing,mayer2150benefits} built on frameworks such as Solid \cite{sambra2016solid,van2024community}, users maintain ownership and control over their data through personal online data stores (pods), which are hosted on Solid-compliant servers~\cite{mansour2016demonstration,vandenbrande2023pod}. These pods are protected by access control policies tied to WebID-based authentication~\footnote{Solid Protocol Authentication: \url{https://solidproject.org/TR/protocol\#authentication}}, allowing fine-grained permissions on what data can be accessed and by whom.  
ESPRESSO~\footnote{\url{https://espressoproject.org/}}~\cite{ragab2024espresso} presents a decentralized search architecture that enables secure, keyword-based querying across distributed pods without centralizing user data. Queries return only references (i.e., resource URLs) to resources within pods that the querying WebID is authorized to access. ESPRESSO ensures that neither raw resources nor search metadata are leaked to unauthorized parties during search \cite{ragab2024decentralized,bahrani2026rethinking,icwe,ragab2024unlocking}.
However, enabling search functionality over private and distributed data introduces new privacy challenges~\cite{esposito2023assessing,gil2026security}: adversaries may attempt to infer sensitive information from access patterns, index structures, or result statistics, even when access control is respected for the files themselves.
To this end, this paper defines the threat model for ESPRESSO, in which we characterize the types of adversaries considered, describe their capabilities and goals, outline system assumptions, and present the privacy guarantees our system provides to defend against them. 

This paper is structured are as follows. Section~\ref{sec:architecture} provides an overview of the ESPRESSO environment and describes the data storage model and data visibility constraints that shape the decentralized search and recommendation~\cite{moawad2019minaret}. Section~\ref{sec:EspressoSearch} describes how search is carried out in ESPRESSO, including architecture, metadata generation and management for source selection and ranking. The workflow of ESPRESSO including indexing, pod registration for search, metadata management and query processing is presented in Section~\ref{howespressoworks}. Section~\ref{sec:threat-model} presents the threat model, analyzes potential privacy and security risks and discusses design implications. Section~\ref{sec:scope-limitations} describe the scope and limitations and Section~\ref{sec:conclusion-future} concludes the paper.

\section{ESPRESSO Overview and Scope} \label{sec:architecture}

Before we delve into the trust assumptions and the details of our threat model, we first introduce the ESPRESSO framework, its infrastructure, and main components.
%
ESPRESSO focuses on enabling efficient discovery of resources that are both relevant to a search party’s query and permitted under their access rights. A search party may be a human user or an automated agent or bot. ESPRESSO is deployed within the \textit{Solid} ecosystem, where users store data in personal online data stores (or pods) hosted on Solid-compliant \footnote{Solid Protocol: Draft Community Group Report, 23 April 2026 \url{https://solidproject.org/TR/protocol}, last accessed 23 April 2026} servers. Each pod enforces access control using the Web Access Control (WAC) specification~\footnote{Web Access Control: Draft Community Group Report, 23 April 2026 \url{https://solidproject.org/TR/wac}, last accessed 23 April 2026} and associates permissions with authenticated WebIDs. 

\subsection{Data Stores and Data Visibility scopes}

In ESPRESSO, each \textit{Data Store} (Solid pod) \( P_i \) is owned by a data owner \( O_i \) and contains \textit{Resources} \( R_i = \{ r_{i1}, \dots, r_{in_i} \} \). Each item has a simple \textit{access control policy} specifying which search parties (identified by \textit{Universal Unique Identifiers} (UUIDs)), also known as \textit{WebIDs}, can access it. 
These access specifications define a \textit{visibility scope} \( \mathcal{V}_i(U) \), representing the subset of \( P_i \)'s resources visible to a given search party \( U \). 

For example, if \( P_1 \) contains \( \{r_1, r_2, r_3\} \) with access rules: \( r_1 \) is visible to \( \text{UUID}_1 \),  \( r_2 \) to \( \text{UUID}_2 \), \( r_3 \) to \( \text{UUID}_1 \) and \( \text{UUID}_3 \), then \( \text{UUID}_1 \) can access \( \{r_1, r_3\} \) and \( \text{UUID}_2 \) can access \( \{r_2\} \), while \( \text{UUID}_4 \) can view nothing. Extending this across multiple data stores, the \textit{global visibility} of a search party \( U \) is the union of its visibility scopes across a specific number (n) of data stores (i.e.,  \( V(U) = \bigcup_{i=1}^{n} \mathcal{V}_i(U) \)).


ESPRESSO leverages this infrastructure to determine visibility scopes and to enforce that all search operations are scoped to the user’s permissions and conducted without moving raw data out of the pods. 

\section{Search within ESPRESSO} \label{sec:EspressoSearch}
Given a keyword-based query \( Q \) from a search party \( U \), the goal of ESPRESSO is to retrieve matching resources across distributed data stores \( \{P_1, \dots, P_n\} \) while enforcing their data visibility constraints \( \mathcal{V}(U) \) over those stores.  
For a multi-keyword query \( Q = \{ kwd_1, \dots, kwd_m \} \), the result set is:  
\[
\mathcal{R} (Q, U ) = \bigcup_{i=1}^{n} \{ r \in R_i \mid r \text{ ~contains~ } ~kwd_j~ \forall ~kwd_j~ \in Q, \, d \in \mathcal{V}_i(U) \}
\]
\\

ESPRESSO aims to balance retrieval performance with privacy guarantees (please see Section~\ref{privacy-guarntees}). It ensures that query execution respects a search party's access boundaries: an unauthorized search party’s visibility scope excludes restricted data, so it cannot view it; likewise, an authorized search party’s queries are evaluated only over the data within its visibility scope, and cannot reach beyond it. 

\begin{figure}[t]
    \centering	
    \includegraphics[width=\linewidth]{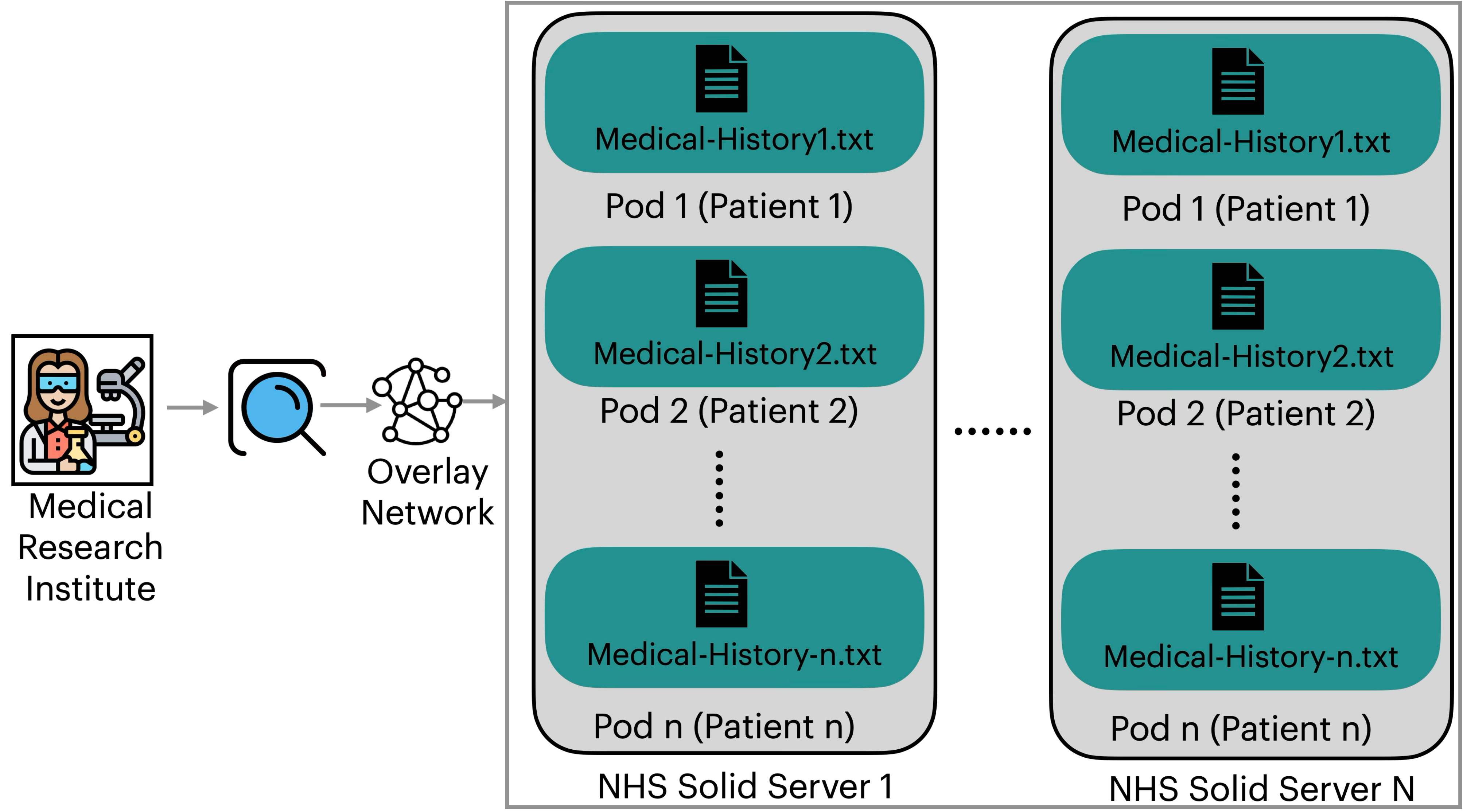}
    \caption{Our Motivating Search Scenario}
	\label{fig:network}
\end{figure}

\paragraph{\textbf{Motivating Scenario}}\label{sec:motiv}
To make our approach more concrete, Figure~\ref{fig:network} depicts a hypothetical healthcare scenario inspired by the organizational structure of the United Kingdom's healthcare ecosystem. In this scenario, we assume a \texttt{Solid}-based environment in which individuals store their medical data in NHS-hosted pods. Patients may upload their medical histories into these pods, including General Practitioner (\texttt{GP}) notes, hospital visit summaries, and related clinical documentation. Consider Alice, a medical researcher seeking to identify potential participants for an authorized early-stage clinical study involving a novel therapeutic intervention. Alice aims to search patient records to identify suitable candidates. To do so, Alice must first obtain permission from individuals to access their pods and, consequently, the medical data contained therein. She intends to issue keyword-based queries describing rare conditions or specific medications across the pods of those who have granted her access.

This scenario necessitates addressing three central questions:
\begin{itemize}

\item \textit{How to enable search that is both efficient and of high quality (ranking) despite varying granular access constraints and the need to operate across multiple infrastructural layers, ranging from text resources to pods to servers?}

\item \textit{How can decentralized indexing enable authorized search parties, like Alice, to undertake privacy-preserving keyword searches across large numbers of data stores with heterogeneous access control constraints?}

\item \textit{How can search efficiency scale with growing decentralized data ecosystems while preserving privacy? For instance, how efficient would Alice’s search be across hundreds of thousands of patient data stores in a metropolitan area?}



\end{itemize}

To enable this workflow, individuals must first approve Alice's access request by adding her WebID to the Access Control Lists associated with the medical records they intend to share. Once authorised, Alice may use her WebID to execute searches across the pods to which access has been granted, allowing her to identify individuals whose records contain the targeted conditions or medications. Moreover, individuals who have chosen to make their contact details accessible to Alice's WebID can subsequently be approached directly, enabling Alice to invite them to participate in the clinical trial.

\begin{figure*}[t]
	\centering
   \includegraphics[width=0.75\linewidth]{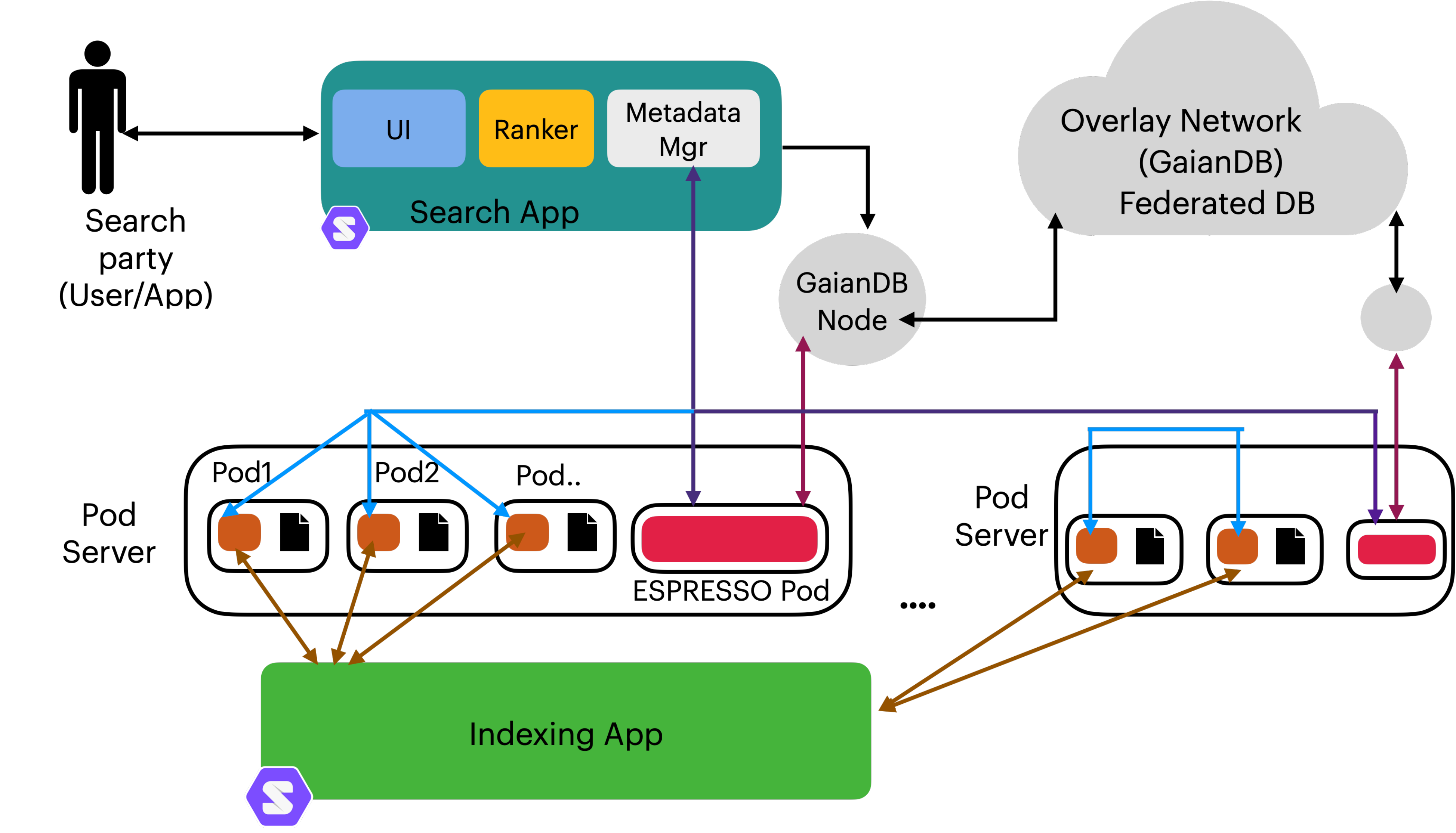}
	\caption{ESPRESSO Framework Architecture}
	\label{fig:DecentSearch}
\end{figure*}

\subsection{ESPRESSO Infrastructure and Components}

The ESPRESSO architecture (Figure~\ref{fig:DecentSearch}) comprises the following components:
\begin{itemize}
    \item \textbf{Indexing App:} This is a Solid-compliant app which creates search-party-scoped indexes based on pod content that the search party is allowed to access and contributes metadata for aggregation into Server-level metadata, acting on behalf of and with the credentials of the pod owner;
    \item \textbf{Search App:} This is another Solid-compliant app which verifies search party's credentials and answers their queries by consulting the appropriate scoped index on behalf of and with the credentials of a search party and of the ESPRESSO system;
    \item \textbf{A Metadata Manager}, a component responsible for maintaining and updating metadata used for source selection and decentralized results ranking by processing and aggregating metadata contributed by the indexing app of pod owners. It acts on behalf of and with the credentials of the ESPRESSO system. 

    \item \textbf{Overlay Network} ESPRESSO also utilizes an Overlay Network in order to connect and catalogue multiple Solid servers, building up a community of connected servers. Specifically, we utilize \textit{GaianDB}~\footnote{Choosing IBM GaianDB (\url{https://github.com/gaiandb/gaiandb}) is an implementation choice rather than a requirement of the framework: in principle, the same metadata could also be maintained in a centralized store or a blockchain-based layer.}, as a federated storage layer for metadata about the Solid servers. ESPRESSO uses \textit{Logical Tables} that works as an abstract federation layer within the GaianDB network, integrating data from different Solid servers. In this sense, GaianDB is not a mere P2P overlay network; it also serves as a distributed data federation platform that enables data to be stored and retrieved from multiple locations, adhering to the principle of \textit{Store Locally and Query Anywhere}.

\end{itemize}


\subsubsection{\textbf{Decentralized Indexing}} \label{subsec:indexes}
ESPRESSO provides an Indexing App that maintains local inverted indexes within each Solid pod. Unlike traditional search systems, ESPRESSO does not aggregate indexes in a centralized location. Instead, the Indexing app generates and maintains \textbf{search party-specific index} files within each pod. Each search party (identified by a \textit{WebID}, which is the UUID in the Web and Solid contexts) has a dedicated index file that contains reference only to the data they have been granted the right to access. In particular, each WebID-specific index file contains an inverted index for the keywords/terms appearing in the resources accessible to that WebID within this pod. In addition, a separate index file is created for all the resources within the pod that have been designated by its owner as being "publicly accessible". This ensures that each search party retrieves results \textit{only} comprising resources they have explicit access to, enforcing data access rights.

Metadata generation and indexing are not \textit{one-off} operations. In practice, both pod-level indexes, the derived server-level and system-level metadata must be maintained and updated over time to reflect changes in pod contents, resource deletions, newly added resources, and modifications to access control constraints. In the current ESPRESSO prototype, this maintenance is not yet fully automated; however, any practical deployment of ESPRESSO would require periodic or event-driven re-indexing and metadata refresh to preserve both retrieval correctness and privacy guarantees.

A further practical dependency is that indexing requires appropriate authorization in addition to resource-level sharing. If a pod owner grants a search party access to a resource but does not grant the ESPRESSO Indexing App sufficient permission to read that resource and update the corresponding index and metadata profile, then that resource may remain \textit{undiscoverable} through ESPRESSO even though its underlying document permissions would otherwise allow access. Thus, searchable participation in ESPRESSO requires not only granting access by data owners to other parties at the resource level, but also the pod owner's explicit cooperation in enabling indexing and metadata maintenance.

\subsubsection{\textbf{Source Selection Metadata Generation}}
In ESPRESSO, metadata generation occurs in two phases: the Indexing App and the Metadata Manager. During the Indexing stage, the Indexing App, crawls text resources within each pod and also checks the Access Control List (ACL) associated with each resource (subject to the pod owner's permission). Based on accessible content, it creates WebID-specific inverted indexes for resources to which a WebID has been added specifically to an ACL along with a public index for \textit{openly accessible} data within each pod. The indexes contain keywords and references to the corresponding resource (URLs). As part of the same process, the Indexing App creates a metadata profile summarizing the content of each indexed pod. The Indexing App then stores this metadata profile in the ESPRESSO pod on the hosting server, which serves as a central location for maintaining the metadata associated with all the pods hosted on this server.

When a pod is registered for search, the pod owner grants the Search App WebID permission to access the metadata profile associated with the pod. This permission applies only to the metadata profile stored by the Indexing App in the ESPRESSO pod on the hosting server, not the related pod resources. The envisaged Metadata Manager uses the same Search App WebID to access the metadata profile. It periodically retrieves these stored metadata profiles from ESPRESSO pod and aggregates them to form the higher-level metadata structures, at the server level and at the system (overlay network) level. On each server, this process results in Server-Level metadata that captures keywords, pod URLs, and Server-Level statistics. This derived metadata is stored inside the ESPRESSO Pod. At the overlay network level, the System-Level metadata includes keywords, server addresses, and network-level statistics. The metadata is maintained in a WebID-scoped manner at both the Server-Level and the System-Level: for each search party, ESPRESSO derives metadata only from the resources that are visible to that WebID and keeps the resulting metadata logically separated from that of other WebIDs. The Search App later uses this metadata to identify relevant servers and pods during query processing.

\subsubsection{\textbf{Source Selection Metadata Structures}}~\label{subsec:metadata} The Server-level and System-level metadata facilitates distributed query processing by identifying the hosting servers and associated pods most relevant to a given query, and accessible to the search party. It further enables ranking of these candidate sources based on their estimated relevance and the search party’s authorized visibility scope.

\begin{figure}[t]
\centering
\includegraphics[width=\linewidth]{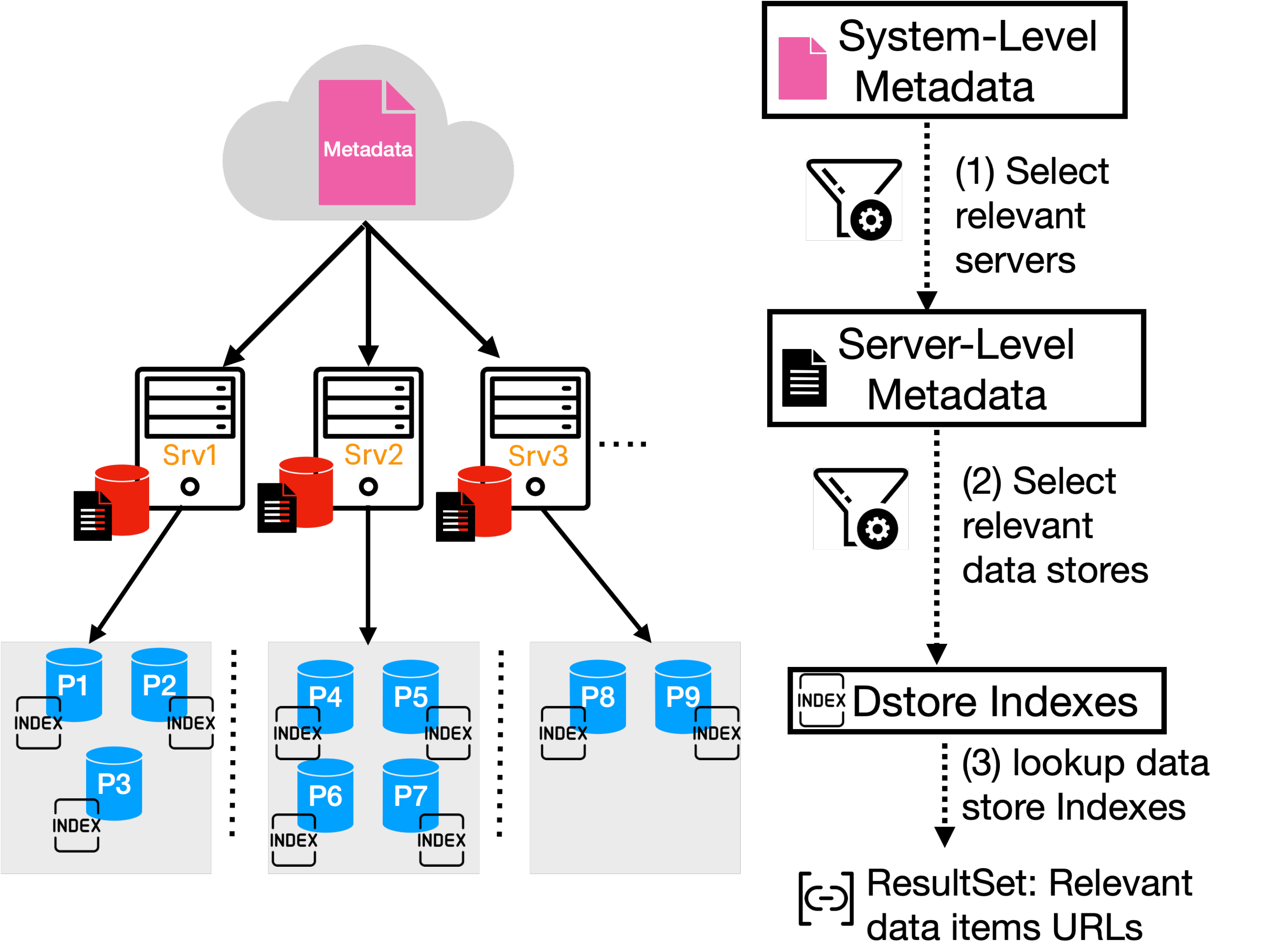}
\caption{ Metadata maintained for source selection. }
\label{fig:metadata}
\end{figure}

Figure~\ref{fig:metadata} illustrates the system-level and server-level metadata maintained by the Metadata Manager in support of source selection.

\begin{itemize}
  
    \item \textbf{System-Level Metadata (metadata about servers hosting data stores)}: A high-level index that includes term distribution statistics (e.g., \textit{term frequency}, \textit{collection size}) and search party visibility (according to UUIDs) across servers. This metadata enables efficient query processing by identifying and ranking relevant servers for a query issued by a specific search party (Step 1 Figure~\ref{fig:metadata}).

Formally, the system-level metadata is a mapping:
\[
MS_N : \mathcal{U} \rightarrow (\mathcal{M} \rightarrow \mathcal{R}_N),
\]
where
\[
\mathcal{R}_N \triangleq \mathcal{P}(\text{Servers}) \times \mathsf{Stats}_N,
\]
and $\mathsf{Stats}_N$ denotes the space of overlay network-level summary statistics.

For each search party $\text{UUID}_j \in \mathcal{U}$ and keyword $kwd_i \in \mathcal{M}$, we define
\[
MS_N(\text{UUID}_j)(kwd_i) \triangleq \langle \mathsf{Srv}_{i,j},\ \mathsf{Stats}^N_{i,j} \rangle.
\]

Here:
\begin{itemize}
    \item $\mathsf{Srv}_{i,j}$ is the set of server addresses (URLs/IDs) that host at least one
    data store containing resources with keyword $kwd_i$ and that are visible to the search party
    identified by $\text{UUID}_j$.
    
    \item $\mathsf{Stats}^N_{i,j}$ stores aggregated network-level distribution statistics for keyword
    $kwd_i$ within the visibility scope of $\text{UUID}_j$, 
    (e.g., number of accessible servers to $\text{UUID}_j$, aggregated term-frequency summaries, aggregated
    collection-size summaries, etc.).
\end{itemize}

If no such servers exist, then $\mathsf{Srv}_{i,j}=\emptyset$ and $\mathsf{Stats}^N_{i,j}$ is empty or zero-valued.

\textbf{\textit{Example:}} For $K=\{kwd_1,kwd_2\}$ and $U=\{\text{UUID}_1,\text{UUID}_2\}$, an
example of $MS_N$ is:
\[
MS_N(\text{UUID}_1)=
\begin{cases}
kwd_1 \mapsto \langle \{\text{server1}\},\ \mathsf{Stats}^N_{1,1}\rangle\\
kwd_2 \mapsto \langle \{\text{server1},\text{server2}\},\ \mathsf{Stats}^N_{2,1}\rangle
\end{cases}
\]

\[
MS_N(\text{UUID}_2)=
\begin{cases}
kwd_1 \mapsto \langle \emptyset,\ \mathbf{0}\rangle\\
kwd_2 \mapsto \langle \{\text{server3}\},\ \mathsf{Stats}^N_{2,2}\rangle
\end{cases}
\]

In this example, we, for simplicity, instantiate $\mathsf{Stats}^N_{i,j}$ with only the cardinality
$|\mathsf{Srv}_{i,j}|$, but the Metadata record supports any other additional/required system-level statistics.

 \noindent
\begin{itemize}
    \item For $kwd_1$ and $\text{UUID}_1$, there is one visible server hosting resources containing $kwd_1$:
    $\mathsf{Srv}_{1,1} = \{\text{server1}\}$ and $\mathcal{S}^N_{1,1} = 1$.
    
    \item For $kwd_2$ and $\text{UUID}_1$, there are two visible servers hosting resources containing $kwd_2$:
    $\mathsf{Srv}_{2,1} = \{\text{server1}, \text{server2}\}$ and $\mathcal{S}^N_{2,1} = 2$.
    
    \item For $kwd_2$ and $\text{UUID}_2$, there is one visible server hosting resources containing $kwd_2$:
    $\mathsf{Srv}_{2,2} = \{\text{server3}\}$ and $\mathcal{S}^N_{2,2} = 1$.
    
    \item For $kwd_1$ and $\text{UUID}_2$, no relevant servers exist:
    $\mathsf{Srv}_{1,2} = \emptyset$ and $\mathcal{S}^N_{1,2} = 0$.
\end{itemize}

\item \textbf{Server-Level Metadata (metadata about data stores hosted on a server)}: Each server hosts metadata about its own pods. This includes keyword distribution, term statistics, and visibility scopes of search parties (UUIDs) across the hosted data stores. It ensures queries only propagate to the relevant pods hosted at that server (Step 2 Figure~\ref{fig:metadata}).

Throughout the paper, we identify each pod (data store) by its URL. Accordingly, we use the terms \emph{pod}, \emph{data store}, and \emph{pod URL} interchangeably, and represent pods directly by their URLs in all metadata structures and result sets.

Similarly to system-level metadata, server-level metadata is scoped by search-party visibility.

Formally, for each server, the server-level metadata is a mapping:
\[
MS_S : \mathcal{U} \rightarrow (\mathcal{M} \rightarrow \mathcal{R}_S),
\]
where
\[
\mathcal{R}_S \triangleq \mathcal{P}(\text{Pods}) \times \mathsf{Stats}^S,
\]
and $\text{Pods}$ denotes the set of pod URLs hosted on server, while $\mathsf{Stats}^S$ denotes
the space of server-level summary statistics.

For each search party $\text{UUID}_j \in \mathcal{U}$ and keyword $kwd_i \in \mathcal{M}$, we define
\[
MS_S(\text{UUID}_j)(kwd_i) \triangleq \langle \mathsf{Pods}_{i,j},\ \mathsf{Stats}^S_{i,j} \rangle.
\]

Here:
\begin{itemize}
    \item $\mathsf{Pods}_{i,j}$ is the set of pod (data store) URLs hosted on server $s$ that contain
    resources with keyword $kwd_i$ and that are visible to the search party identified by $\text{UUID}_j$.

    \item $\mathsf{Stats}^S_{i,j}$ stores server-level distribution statistics for keyword $kwd_i$ within
    the visibility scope of $\text{UUID}_j$, used for pod selection and ranking (e.g., number of visible pods,
    aggregated term-frequency summaries, collection-size summaries, etc.).
\end{itemize}

If no such pods exist, then $\mathsf{Pods}_{i,j}=\emptyset$ and $\mathsf{Stats}^S_{i,j}$ is empty
or zero-valued.

\textbf{\textit{Example:}} For $K=\{kwd_1,kwd_2\}$ and $U=\{\text{UUID}_1,\text{UUID}_2\}$, an example of
$MS_S$ is:
\[
MS_S(\text{UUID}_1)=
\begin{cases}
kwd_1 \mapsto \langle \{\text{dstore1}\},\ \mathsf{Stats}^S_{1,1}\rangle\\
kwd_2 \mapsto \langle \{\text{dstore2},\text{dstore3}\},\ \mathsf{Stats}^S_{2,1}\rangle
\end{cases}
\]

\[
MS_S(\text{UUID}_2)=
\begin{cases}
kwd_1 \mapsto \langle \emptyset,\ \mathbf{0}\rangle\\
kwd_2 \mapsto \langle \{\text{dstore4}\},\ \mathsf{Stats}^S_{2,2}\rangle
\end{cases}
\]

In this example, we instantiate $\mathsf{Stats}^S_{i,j}$ with only the cardinality
$|\mathsf{Pods}_{i,j}|$, but similar to the system-level metadata, the record supports any Server-level statistics required.

\noindent
\begin{itemize}
    \item For $kwd_1$ and $\text{UUID}_1$, there is one visible pod hosting resources containing $kwd_1$:
    $\mathsf{Pods}_{1,1} = \{\text{dstore1}\}$ and $\mathsf{Stats}^S_{1,1} = 1$.
    
    \item For $kwd_2$ and $\text{UUID}_1$, there are two visible pods hosting resources containing $kwd_2$:
    $\mathsf{Pods}_{2,1} = \{\text{dstore2}, \text{dstore3}\}$ and $\mathsf{Stats}^S_{2,1} = 2$.
    
    \item For $kwd_2$ and $\text{UUID}_2$, there is one visible pod hosting resources containing $kwd_2$:
    $\mathsf{Pods}_{2,2} = \{\text{dstore4}\}$ and $\mathsf{Stats}^S_{2,2} = 1$.
    
    \item For $kwd_1$ and $\text{UUID}_2$, no relevant pods exist:
    $\mathsf{Pods}_{1,2} = \emptyset$ and $\mathsf{Stats}^S_{1,2} = 0$.
\end{itemize}

    \item \textbf{Probabilistic Source Selection Metadata}: 
    Source-selection metadata can leverage inverted indexes to provide precise mappings to relevant sources (servers and data stores). However, our work also contrasts this representation with {\em probabilistic} metadata structures to evaluate their effectiveness and efficiency trade-offs during decentralized query processing. These latter techniques represent metadata at various levels without exposing precise data store or server information. To achieve this, we apply a Bloom filter mechanism over the search party-specific metadata structures and indexes for ACL-protected resources, while publicly accessible resources are represented separately through shared aggregated metadata that is common to all search parties. These probabilistic structures encode potential keyword matches associated with search parties' UUIDs while preventing exact reconstruction of the metadata. Additionally, since each Bloom filter is search party-specific, access is restricted to the search party possessing the corresponding UUID, thereby reinforcing query privacy.

\end{itemize}

After identifying relevant servers and data stores, the system navigates only to the selected servers and queries only the selected pod indexes, aggregates results, and returns them to the search party (Step 3 Figure~\ref{fig:metadata}). This ensures privacy preservation while optimizing query performance. 

\section{ESPRESSO Workflow}~\label{howespressoworks}

In this section, we provide details on how the main components of ESPRESSO work together to enable decentralized search over personal data stores distributed across multiple Solid servers. This discussion lays the groundwork for understanding how the ESPRESSO experimentation environment and its components are well-suited to supporting a decentralized search solution, as well as how different stakeholders could effectively engage with it.
\begin{figure}[t]
    \centering
    \includegraphics[width=0.5\textwidth]{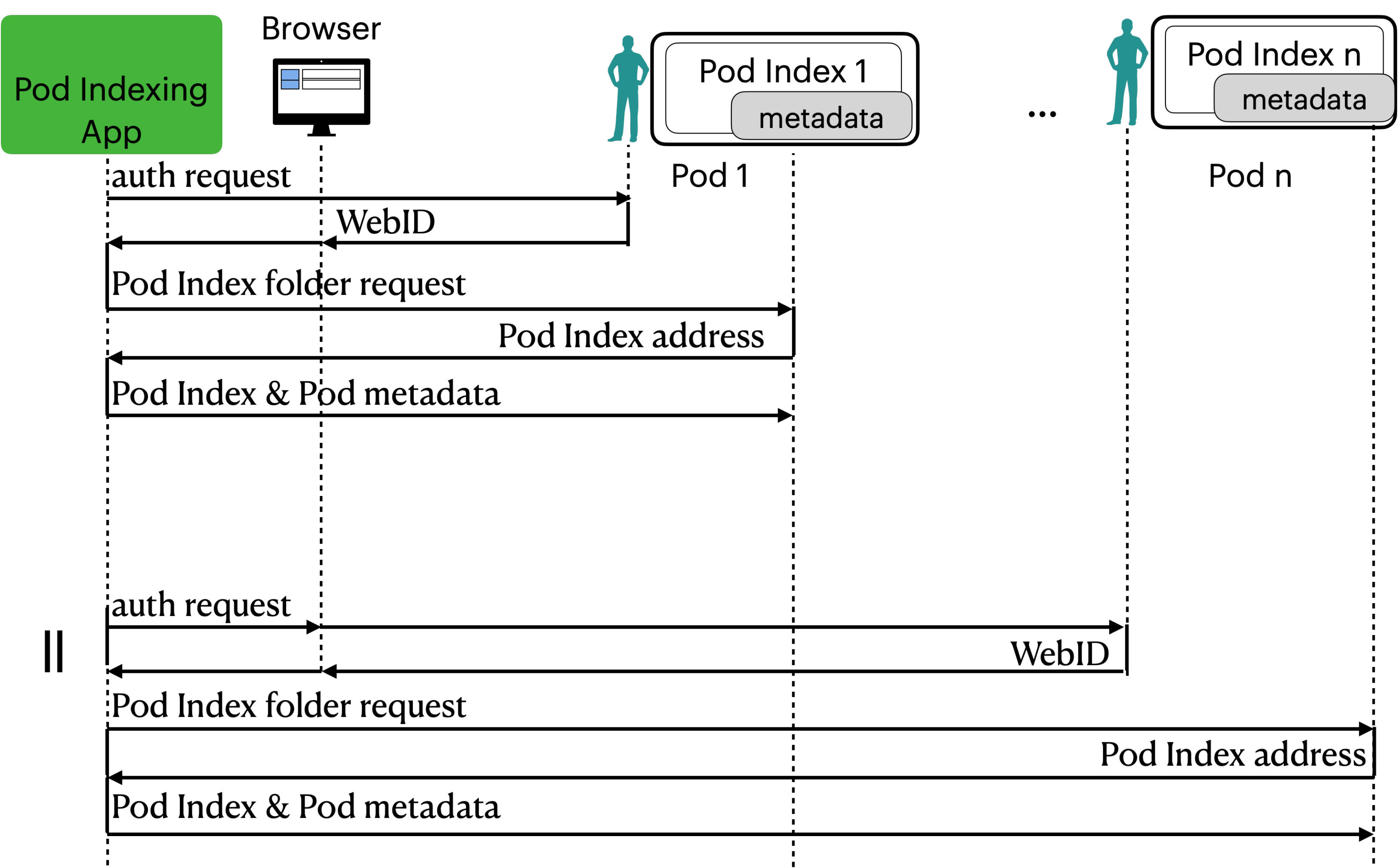}
    \caption{Indexing Process in ESPRESSO}
    \label{fig:indexingsequence}
\end{figure}

\subsection{Pods Indexing}

The \textit{Indexing App} manages the indexing of unstructured text data in pods. Figure~\ref{fig:indexingsequence} shows a sequence of interactions between the Indexing App, the owner of the pod, and the target pod. The process begins when the Indexing App initiates an authentication request through the browser interface. The pod owner logs in using their WebID credentials and authorizes the Indexing App to access the pod. After initial authorization, the Indexing App obtains an access token that allows it to perform subsequent indexing operations without requiring repeated user authentication until the token expires. Once authenticated, the Indexing App accesses a well-known index directory within the pod, which is standardized across all pods participating in the ESPRESSO system. The Indexing App processes the pod’s data and creates inverted index files locally within the pod along with the metadata profile for the ESPRESSO server pod is to support keyword-based search. The indexing process is repeated independently for each pod participating in the ESPRESSO system. The generated indexes cover both publicly accessible resources and the resources accessible to specific Web-ID based on ACL permissions.

\begin{figure}[h]
    \centering
    \includegraphics[width=0.5\textwidth]{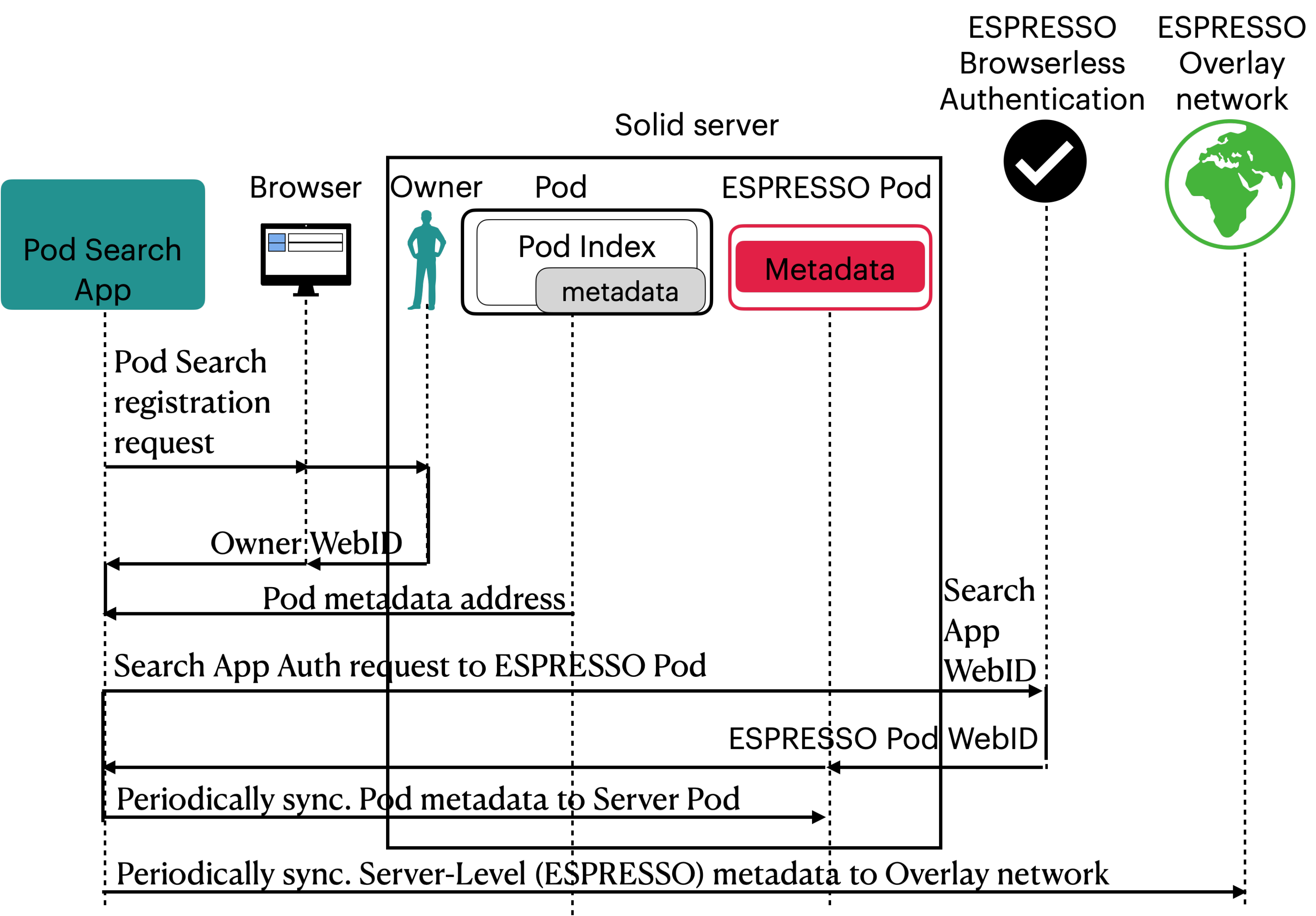}
    \caption{The process of registering a Pod for Search}
    \label{fig:registrationsequence}
\end{figure}

\subsection{Servers Registration for Search}~\label{appendixsubsec:podregisteration}
Before individual pods can participate in ESPRESSO search, the Solid server must first be prepared to host the ESPRESSO metadata structures. This server-level registration step establishes the ESPRESSO pod on the server, which acts as a dedicated system-managed location for storing metadata associated with searchable pods hosted on that server. Concretely, the Search App WebID (or ESPRESSO system WebID) is used to create the ESPRESSO pod within the Solid server. This pod is distinct from user pods and is reserved for ESPRESSO-related metadata and coordination functions.
Within this ESPRESSO pod, the server maintains a directory structure through which metadata associated with registered pods can be stored and accessed in a controlled manner. This step is a prerequisite for pod-level registration, since it provides the server-side location in which pod metadata profiles will later be deposited and maintained. In this way, ESPRESSO first establishes the shared metadata container at the server level, and only then enables individual pod owners to register their pods for participation in decentralized search.

\subsection{Pods Registration for Search}~\label{appendixsubsec:podregisteration}
After the ESPRESSO pod has been created on the Solid server, individual pod owners may register their indexed pods for search participation. 
%
Figure~\ref{fig:registrationsequence} shows a sequence of steps between the Pod Search App, the pod owner, and the Solid Server environment. The process is initiated when the pod owner submits a registration request through the Search App interface. The owner then authenticates using their WebID credentials to authorize the Search App WebID to access the metadata profile associated with the pod. The pod metadata profile is stored at a well-known location in the ESPRESSO pod hierarchy based on the WebID of the pod owner. Following the authorization, the Search App obtains an access token that allows it to access the ESPRESSO Pod metadata for further operations until the token expires. This token can be safely stored and reused without requiring frequent user authentication. The access control rules of the metadata profile are updated to allow it to be used by the ESPRESSO search system. Once the registration is completed, the pod can participate in decentralized search while still enforcing the access permissions set by the pod owner. 

\subsection{Source Selection Metadata Flow} ~\label{appendixsubsec:metadatmanger}
To support the efficient search across multiple servers, ESPRESSO envisages a Metadata Manager that is responsible for maintaining the up-to-date metadata. This component is not yet implemented in the ESPRESSO experimental environment. While the generation of metadata is described earlier, this section explains how the metadata is periodically synchronized and used during the system operation. The Metadata Manager regularly retrieves metadata associated with each server pod and consolidates it so that an updated view of searchable pods can be maintained locally. This integrated metadata is stored within the ESPRESSO Pod. In addition, on each server the metadata manager shares the summary of server-level metadata with the overlay network, so that relevant servers can be identified during search. The information about the higher-level metadata is provided in the Section~\ref{subsec:metadata}.

The Metadata Manager will use the Search App's authorized WebID to periodically access and copy those servers' meta-information from ESPRESSO Pods at each server and store the locations of the Server-Level Metadata in the logical tables of the Overlay Network. Notably, metadata maintained in the Overlay Network logical tables will not reveal any sensitive information to unauthorized parties, as the tables relate the WebIDs to the ESPRESSO Pod URLs that contain Server-Level Metadata information accessible to those WebIDs. The ESPRESSO Pod URLs can only be accessed by the ESPRESSO Search App WebID.

\subsection{Search Pipeline within the Search App.}~\label{appendixsubsec:searchapp}
Search parties initiate searches through a dedicated User Interface (UI), submitting a search request (multi-keyword query) to the Search App along with a WebID. The Search App processes these requests by utilizing the Overlay Network to perform searches across Solid servers, ensuring that results include only references to resources accessible to the Search Party's WebID.

\begin{figure}[h]
    \centering
    \includegraphics[width=0.5\textwidth]{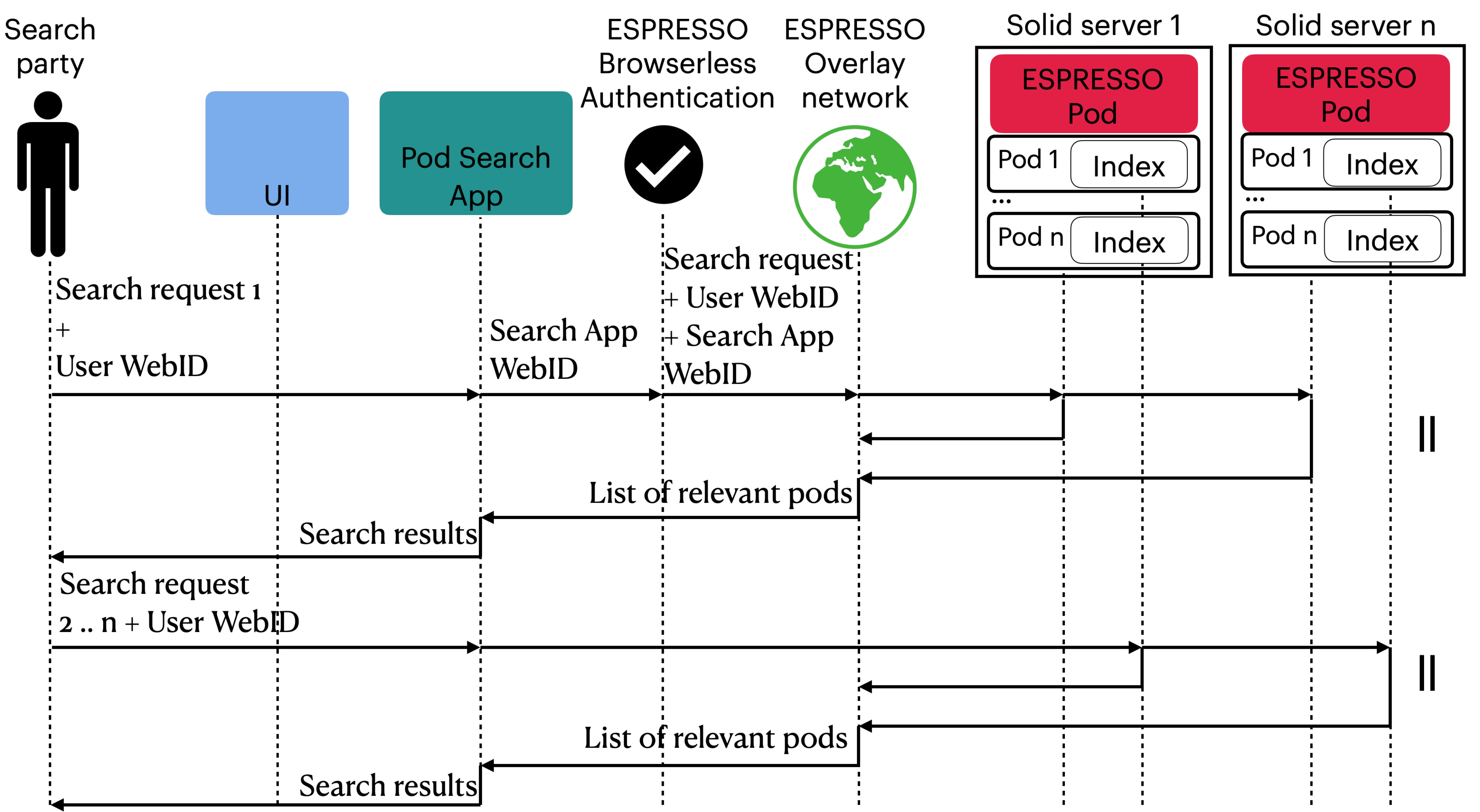}
    \caption{Search Process when overlay returns relevant pods}
    \label{fig:search1}
\end{figure}

\begin{figure}[h]
    \centering
    \includegraphics[width=0.5\textwidth]{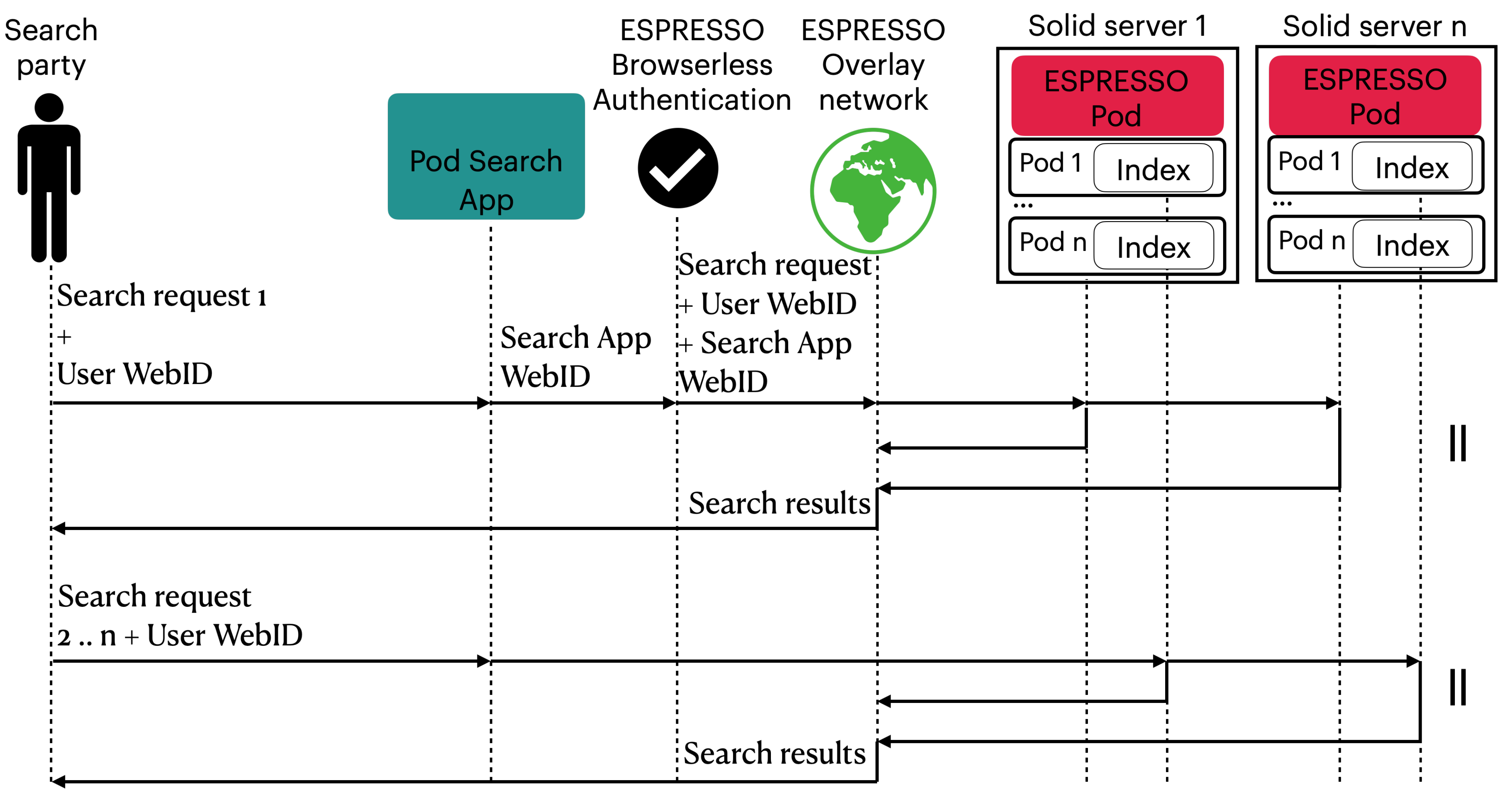}
    \caption{Search Process when overlay returns results}
    \label{fig:search2}
\end{figure}

The Search App employs two primary alternative methods to utilize the support of the GaianDB network. 

\begin{itemize}
    \item  In the first approach, the Search App utilizes GaianDB as a distributed storage layer for storing the System-Level Metadata about servers, maintained across GaianDB nodes within the Overlay Network logical tables. Thus, the Overlay Network responds by providing a curated list of relevant servers to the Search App. Subsequently, the Search App accesses the Server-Level Metadata about pods, through their respective servers, to get the relevant pods. For each relevant pod, the Search App will access the WebID-specific Pod Index, and get the results accessible by that WebID. Thus, the Search App needs to conduct subsequent HTTP GET requests to the relevant Pod Index folders, utilizing permissions granted to the search party WebID. This process allows the Search App to retrieve results based on the specified search party WebID, facilitating efficient information retrieval across the network.

    \item For the second search approach, the Search App leverages the Overlay Network to transmit both the query and the WebID to the GaianDB node. From there, the GaianDB node proceeds to distribute and propagate these requests across the Overlay Network nodes. The GaianDB Overlay Network then aggregates the results of these queries and returns them directly to the Search App user interface. This method streamlines the process of querying and retrieving information by centralizing the aggregation of results through the GaianDB node, enhancing efficiency in data retrieval for the Search App.

\end{itemize}

\section{Threat Model \& Privacy Guarantees} \label{sec:threat-model}
In this section, we present the adversarial assumptions and formal privacy guarantees underpinning the decentralized search within our ESPRESSO framework. We leverage Microsoft's threat modeling \textit{STRIDE} framework\footnote{STRIDE THREAT Modelling: \url{https://learn.microsoft.com/en-us/azure/security/develop/threat-modeling-tool-threats}}. We have chosen STRIDE for its comprehensive approach and suitability for a distributed system consisting of multiple components of different trust levels.

While STRIDE encompasses a broad spectrum of system-security concerns, we focus selectively on the threat categories that are most relevant to our setting. For each relevant threat, we describe the corresponding adversarial capabilities and explain how the design of ESPRESSO mitigates or eliminates the associated risks. More specifically, from STRIDE, we leverage only the threats most relevant to our ESPRESSO decentralized search framework, namely, \textit{\textbf{Information Disclosure}} and \textit{\textbf{Spoofing}}. 

\begin{enumerate}
    \item \textit{\textbf{Information Disclosure}}: Information is exposed to individuals not authorized to see it, i.e., sensitive data is leaked beyond the intended access boundary.
    \item \textit{\textbf{Spoofing}}: An adversary impersonates another entity, such as a user or a system component.
\end{enumerate}

While the remaining STRIDE categories—Tampering, Repudiation, and Denial of Service (DoS)—are fundamental system-security concerns, they are less central to the specific threat model addressed in ESPRESSO:
\begin{enumerate}
    \item Tampering: ESPRESSO assumes integrity-protected storage within user-controlled pods and secure communication channels. As a result, the risk of unauthorized modification of data or metadata is minimized by underlying platform guarantees rather than by ESPRESSO’s search design itself.
    
    \item Repudiation: ESPRESSO is designed for privacy-preserving decentralized search rather than auditability or forensic traceability. Since ESPRESSO does not rely on user-generated logs or non-repudiable actions for correctness, repudiation threats fall outside the core scope of our analysis.
   
   \item Denial of Service (DoS): Although DoS is a practical concern in any distributed system, addressing availability attacks (e.g., overloaded servers or rate limiting) is largely an operational and deployment-level issue. ESPRESSO’s protocol does not introduce additional DoS vulnerabilities beyond those inherent in typical web-based services, so DoS resilience is considered orthogonal to our design goals.

\end{enumerate}

By narrowing our focus to Information Disclosure and Spoofing, we align our security analysis here with the primary privacy and authenticity concerns that are inherent in decentralized search over personal pods.

\subsection{Adversary Model}
The STRIDE threats of Information Disclosure and Spoofing motivate our adversary model. We distinguish between (i) the \textit{type} of adversary, which captures where the adversary is located in the system, and (ii) the \textit{profile} of the adversary, which captures how the adversary behaves. 

\paragraph{\textbf{Adversary Profiles:}} 

Information Disclosure and Spoofing let us consider both \textbf{honest-but-curious} and \textbf{malicious} adversary profiles. While \textbf{malicious adversaries}  relate to the Spoofing threat and impersonation attacks, the \textbf{Honest-but-curious adversaries} relate to  Information disclosure attacks. 

\begin{itemize}
    \item \textbf{Malicious adversaries (Spoofing)} Malicious adversaries may attempt to impersonate a legitimate search party by exploiting weaknesses in authentication or session management. This includes attacks such as session fixation, token replay, misuse of delegated agents or applications acting on behalf of a WebID. Such attacks aim to submit queries under an unauthorized WebID in order to access data or metadata beyond the adversary’s legitimate visibility scope.

   \item  \textbf{Honest-but-curious adversaries} Those are a direct manifestation of the \textit{Information Disclosure} threat. They are users who follow the protocol using their own authenticated WebID but may attempt to infer additional information from query patterns, result cardinalities, or metadata structures. For instance, a clinician with access to three patients may try to estimate how many other users on the system also mention “diabetes” by analyzing Bloom filters or response sizes.
\end{itemize}

\paragraph{\textbf{Adversary Types:}} We consider the following adversary types, operating independently and without coordination or collusion:

\begin{itemize}
    \item \textbf{Search party} $U$ who issues queries and may attempt to learn information beyond their visibility scope $V(U)$.
    \item \textbf{Hosting servers} $S_k$ that host Solid pods, correctly enforce Access Control Policies and Web Access  Control mechanisms (ACP/WAC) and pod-level index isolation, but may observe query execution and local metadata.
    \item \textbf{Overlay nodes} that participate in query routing or metadata aggregation and observe overlay-level information.
\end{itemize}

In this threat model, we do \emph{not} consider collusion between adversaries (e.g., a user and a server sharing information, or multiple users pooling access). Our guarantees are designed under the assumption that adversaries act individually. Extending the model to account for collusion is a direction for future work.

\paragraph{\textbf{Adversary Capabilities:}} Adversaries in our model may potentially observe or access the following information while following the search protocol, forming the basis for the inference attacks described above:

\begin{itemize}
    \item \textbf{WebID-scoped Index Access:} A search party $U$ may access the pod-level inverted indexes $\mathcal{I}_{ij}$ built from their visibility scope $V_i(U)$.
    \item \textbf{Metadata Exposure:} Adversaries may inspect server-level or overlay network-level metadata, either in a deterministic way (exact counts, URLs) using inverted indexes or in a probabilistic form using Bloom filters.
    \item \textbf{Query Results:} The set of resource references (URLs) and relevance scores returned in response to queries submitted by search parties.
    \item \textbf{Query Execution Side-channels:} Timing information, network traffic patterns, and which servers or pods are contacted during query processing.
\end{itemize}

\paragraph{\textbf{Adversaries' Goals:}}

The adversarial goals we consider primarily fall under the \textit{Information Disclosure} threat. While \textit{Spoofing} (impersonation) is prevented by Solid’s authentication layer (see Section~\ref{subsec:asuump-limit}), our analysis focuses on inference attacks that remain possible \emph{after} legitimate authentication. We categorize adversarial goals into two broad classes: (1) \emph{General reconstruction and auxiliary knowledge inference}, and (2) \emph{Individual-level identification or re-identification}. Both classes are relevant in the context of health data and Solid pods, where seemingly innocuous metadata may be used to infer sensitive information. 

\medskip

\noindent\textbf{1. General Reconstruction and Inference}

These attacks aim to recover structural or statistical properties of the dataset across pods, potentially revealing sensitive population-level or organizational information.

\begin{enumerate}[label=(G\arabic*)]
    \item \emph{Membership Inference:} Determine whether a keyword $t$ or a resource $r$ exists in a pod outside the adversary’s authorized visibility scope. In the context of dealing with sensitive data (e.g., health data), this may reveal unintended associations between users or institutions and sensitive conditions.

    \item \emph{Access Pattern Inference:} Infer which pods or servers are accessed in response to a query. For example, in the case of health data, repeated observations within the search results may reveal which users, groups, or institutions are relevant to specific medical topics (e.g., repeated access during ``mental health'' queries).

    \item \emph{Keyword Frequency Estimation:} Estimate how common certain terms are across the network by analyzing metadata or approximate data structures. For example, an adversary may infer that \texttt{``treatment-resistant depression''} appears more frequently in pods hosted by a particular hospital, revealing prevalence patterns or institutional focus that should remain confidential.

    \item \emph{Index Reconstruction:} Partially or fully reconstruct keyword-to-pod mappings or inverted index structures from source-selection metadata, with the aim of gaining structural insight into restricted data.

    \item \emph{Indirect Inference via Correlation:} Infer the presence of a protected or hidden concept (e.g., \texttt{mental illness}) through correlated observable terms (e.g., \texttt{fatigue}, \texttt{withdrawal}) that statistically or semantically co-occur with the hidden concept. Such attacks exploit distributional patterns rather than direct visibility and are particularly relevant under repeated or aggregate observations.
\end{enumerate}

\medskip
\noindent\textbf{2. Individual Identification and Re-identification}

These attacks combine public or auxiliary background knowledge with query results or metadata to associate sensitive information (e.g., health conditions) with specific individuals.

\begin{enumerate}[label=(I\arabic*)]
    \item \emph{Direct Identification:} Determine whether a known individual (e.g., identified by a WebID or public profile) has a resource matching a sensitive keyword such as \texttt{``diabetes''}.

    \item \emph{Quasi-Identifier Re-identification:} Use auxiliary attributes such as age, location, or gender to link an otherwise anonymous or pseudonymous pod to a real-world identity and infer sensitive information.

    \emph{Example:} If an attacker knows that Bob is a 34-year-old male living in Postcode SO17 and observes a pod matching these demographics that responds to a \texttt{``diabetes''} query, they may infer Bob’s condition.
\end{enumerate}

\subsection{ESPRESSO Privacy Guarantees}~\label{privacy-guarntees}

Privacy in ESPRESSO is achieved through the combination of baseline guarantees provided by Solid and additional privacy-preserving mechanisms introduced by the ESPRESSO architecture. While Solid governs authentication, authorization and data control at the pod level, ESPRESSO extends these foundations to enable privacy-preserving federated search without centralisation of data, indexes, or sensitive metadata. This section clarifies which guarantees are inherited from Solid and which are explicitly introduced by ESPRESSO.

\subsection{Baseline Trust and Privacy Assumptions (Solid and ESPRESSO)}
\label{subsec:asuump-limit}

ESPRESSO builds on the security and privacy mechanisms provided by Solid, subject to explicit trust assumptions about the underlying infrastructure and server behavior.

\paragraph{Solid Baseline Security Guarantees.}
Under Solid’s access-control framework, the following baseline properties are assumed to hold:

\begin{enumerate}
    \item \textbf{Pod Data Ownership:} User data remains stored in personal pods, with ownership retained by the data subject. No data leaves a pod unless permitted by its Access Control Policy (ACP/WAC).
    \item \textbf{Authenticated Access:} Authentication is enforced using WebID, ensuring that all search actions are bound to a verified digital identity.
    \item \textbf{Policy-Based Authorization:} Access Control Policies determine which WebIDs are authorized to read each resource or file.
    \item \textbf{Application Sandboxing:} Applications (i.e., Indexing App and Search App) operate only with explicitly granted permissions and cannot access data beyond their authorized scope.
\end{enumerate}

\paragraph{ESPRESSO System Trust Assumptions.}
Building on these guarantees, ESPRESSO assumes the following system-level trust properties:

\begin{itemize}
    \item \textbf{Trusted Authentication:} Solid servers and client libraries correctly enforce WebID-based authentication, and each query is bound to an authenticated session. ESPRESSO does not implement its own authentication or session management mechanisms. Instead, the ESPRESSO Search App relies on Solid’s WebID authentication flow and associated token handling to bind each query to an authenticated identity. Consequently, the correctness of session establishment, token freshness, replay protection, delegated agent authorization, and revocation semantics is assumed to be provided by the underlying Solid infrastructure. 
    \item \textbf{Correct Access Control Enforcement:} Solid servers correctly enforce ACP/WAC policies for all pod resources.
    \item \textbf{Pod Integrity:} Pods are not compromised and do not leak data through unintended side channels.
    \item \textbf{Hosting Server Correctness Trust:} Hosting servers correctly authenticate WebIDs, enforce access control, and maintain pod-level index isolation as specified by the ESPRESSO protocol. 
\end{itemize}

Therefore, the Spoofing threats introduced by malicious adversaries are mitigated via Solid’s authentication flow: the ESPRESSO Search App validates each query against a session-bound token that confirms WebID ownership. Impersonation attacks are thus prevented at the authentication layer.


\subsection{Privacy Guarantees Introduced by ESPRESSO (Architectural Layer)}
ESPRESSO builds on Solid's privacy foundations specified earlier and introduces new guarantees that enable secure decentralized keyword search while preventing leakage of information beyond access rights.

\paragraph{\textbf{(PG1) Visibility/Access Scope Isolation}}
ESPRESSO defines visibility scopes $V_i(U)$ and ensures that all search operations and metadata exposure remain bounded to the subset of resources authorised for the search party. No keyword or resource reference is revealed outside $V(U)$. Thus, search results returned to a search party $U$ are strictly limited to its authorized data visibility scope:
\[
\forall U \in \mathcal{U},\quad R(Q, U) \subseteq V(U)
\]

That is, the result set does not contain any reference to resources beyond the visibility scope defined by the access control policies associated with the search party $U$. This is the foundational defense against unauthorized Information Disclosure.

\paragraph{Example.}
Let the text resources corpus be
\[
R = \{r_1, r_2, r_3\}
\]
and let two users $U_A$ and $U_B$ have visibility scopes defined by access-control policies:
\[
V(U_A) = \{r_1, r_2\}, \qquad V(U_B) = \{r_3\}.
\]

Assume the keyword \emph{``diabetes''} appears only in $r_3$.
If user $U_A$ issues the query
\[
Q = \text{``diabetes''},
\]
ESPRESSO evaluates the search strictly over $V(U_A)$.
Since no resource in $V(U_A)$ contains the keyword, the result is
\[
R(Q, U_A) = \varnothing.
\]

No reference, metadata, or indication of the existence of $r_3$ is revealed in the result of that query ($Q$) submitted by user $U_A$. Thus, the returned result satisfies
\[
R(Q, U_A) \subseteq V(U_A),
\] demonstrating visibility/access scope isolation.

\paragraph{\textbf{(PG2) WebID-Scoped Decentralised Indexing \& Index Isolation}}

Instead of relying on a global index, the ESPRESSO Indexing App maintains within each pod $P_i$ a separate inverted index $\mathcal{I}_{ij}$ for every search party $U$ based on the subset of resources in $P_i$ that $U$ is authorized to access. Formally, the index exposed to $U$ is constructed over their visibility scope:

\[
\mathcal{I}_{ij} = \texttt{Index}(V_i(U))
\]

It is important to note that only the pod owner, and the relevant components of ESPRESSO, have access to the pod index. Access to the pod index is granted by adding the WebID - in this case, the ESPRESSO indexing app's WebID and the ESPRESSO search app's WebID - to the Access Control List (ACL) of the pod index, at the level of the container.  Each index file within the pod index is search party-specific. That is, each search party (WebID) that has access to resources in a given pod has its \textit{own} index file, which \textit{only} contains keyword statistics for resources that are accessible to that WebID. When searching a pod for resources that match a search query, the ESPRESSO search app will \textit{only} look at the index files for the WebID of the search party. 


This prevents cross-user inference about keywords or the existence of private resources. It also ensures the search operations are privacy-preserving at the index level: a search party cannot infer the presence or frequency of keywords in text resources they are not authorized to view.




\vspace{1ex}
\noindent\textbf{Example:}
Consider a pod $P_i$ that contains three text resources:
\begin{itemize}
\item $r_1$ containing the terms \texttt{"genetic"} and \texttt{"therapy"}
\item $r_2$ containing the term \texttt{"cancer"}
\item $r_3$ containing the terms \texttt{"diabetes"} and \texttt{"diet"}
\end{itemize}

Let the access control policies be:
\begin{itemize}
\item $U_1$ is authorized to access only $r_1$
\item $U_2$ is authorized to access only $r_2$ and $r_3$
\end{itemize}

Then:
\begin{align*}
V_i(U_1) &= {r_1}, V_i(U_2)= {r_2, r_3}
\end{align*}

The system generates two distinct indexes:
\begin{itemize}
\item $\mathcal{I}{ij}$ includes keywords: \texttt{genetic}, \texttt{therapy}
\item $\mathcal{I}{ik}$ includes keywords: \texttt{cancer}, \texttt{diabetes}, \texttt{diet}
\end{itemize}

During query processing, each search party can access only the index constructed for its own WebID.

Last but not least, if a hosting server deviates from the protocol, for example, by bypassing ACP enforcement, merging indexes across pods, this privacy guarantee no longer holds; such fully malicious servers fall outside our threat model. Correctness trust is supported by auditable, open-source Solid servers and client-side checks that query results are scoped to authorized pods, but clients cannot verify the absence of additional server-side logging. Accordingly, our guarantees limit what can be inferred from observable protocol outputs rather than arbitrary server misbehavior.

\paragraph{\textbf{(PG3) Using Probabilistic Source Selection Metadata}}
Building on the probabilistic metadata disclosure strategy described in Section~\ref{subsec:metadata}, ESPRESSO provides a concrete privacy guarantee by applying Bloom filters to the search-party-specific metadata maintained at both the overlay network level and the server level. The Bloom filter mechanism probabilistically encodes potential keyword occurrences associated with a search party’s WebID, preventing the exact reconstruction of keyword–pod mappings or the underlying metadata structures. Each Bloom filter is generated and \textbf{scoped per search party}, ensuring that only the holder of the corresponding WebID can access it. This search-party-specific encoding restricts other search parties from inferring relationships between keywords, pods, or queries belonging to a different search party, thereby strengthening the confidentiality of both metadata and data within pods. Our Bloom filter approach makes it difficult for an adversary to infer information, since the false positive rate increases with the number of keywords in a query. False positives introduce additional noise, making it difficult to determine the precise query answer from the top retrieved resources.



\paragraph{\textbf{(PG4) Source-Selection Metadata Conservativity \& Separability.}}

Source-selection metadata structures in ESPRESSO are primarily designed to improve system performance and scalability, particularly in large-scale decentralized deployments (e.g., hundreds of servers hosting thousands of pods). Accordingly, while the primary purpose of source-selection metadata at both levels is efficient query processing, ESPRESSO explicitly incorporates privacy considerations into its design. In particular, source-selection metadata structures are required to be both \emph{conservative} and \emph{separable}. Throughout this section, we use the term \emph{search party} interchangeably
with \emph{WebID}.

\begin{itemize}
    \item \textbf{Conservativity} ensures that no source-selection metadata is maintained at any level (server or overlay network) that could not be recomputed solely from the results of \textbf{legitimate} queries submitted by a given authorized search party. In other words, any metadata maintained by ESPRESSO (such as keyword presence/absence, keyword frequency or distribution across sources (servers or pods), or source cardinality information) can be fully derived/reconstructed by a given search party from the observable search results of authorized query executions and does not introduce additional information beyond what those queries already reveal.

    
    \item \textbf{Separability} guarantees that all ESPRESSO source selection metadata is partitioned into disjoint subsets corresponding to \textbf{each search party’s WebID}. This ensures that no search party learns information beyond what they could derive from the results of their authorized queries. From the total server-level or overlay network-level metadata, only the portion that a given WebID could legitimately infer (from the results of its own queries) is accessed when processing queries submitted by that search party, preventing cross-user information leakage. Metadata separability can be viewed as a lifting of index isolation (PG2) from the level of data stores and indexes to the level of source-selection metadata. Concretely, while PG2 ensures that query evaluation and index access are scoped to a single WebID, PG4 ensures that any metadata derived from such evaluation remains partitioned by WebID and does not introduce cross-WebID information flow.

\end{itemize}

To illustrate the above two guarantees, let $\mathcal{W}$ denote the set of WebIDs in the system.
We write $w_1,\dots,w_u \in \mathcal{W}$ for end-user WebIDs, and
$w_{\text{Indexer}}, w_{\text{SearchApp}} \in \mathcal{W}$ for the WebIDs
of the ESPRESSO Indexing App and Search App, respectively.
These principles operate on solid servers $S_1, \dots, S_v$.
A Solid server $S$ hosts pods $P_1, \dots, P_n$ identified by URLs
$\texttt{URL}_1, \dots, \texttt{URL}_n$, of which only a subset $P_1, \dots, P_m$ are accessible to the ESPRESSO Search App. When a search party $w \in \{w_1, \dots, w_u\} \subseteq \mathcal{W}$ issues a query $q$, the ESPRESSO Search App returns to the local overlay node only the URLs $\{URL_1, \dots, URL_k\}$ of pods $(P_1, \dots, P_k) \subseteq (P_1, \dots, P_m)$ that (i) match the query and (ii) are authorized for $w$. We denote this by $\text{Result}(w, q, S)$.

\noindent\paragraph{\textbf{(A) Server-Level Conservativity.}}  
Each Solid server $S$ hosts a set of pods indexed by the ESPRESSO Indexing App and exposes search capabilities via the ESPRESSO Search App. The server-level metadata $MS_S$ maintained on $S$ is computed entirely from access-controlled query results by applying a deterministic function:
\[
MS_S = f_S\left((w_1, q_1, R_1), \dots, (w_r, q_r, R_r)\right),
\]
where $R_i = \texttt{Result}(w_i, q_i, S)$ is the result returned to search party $w_i$ for query $q_i$. Thus, no metadata stored on $S$ contains information that could not be reconstructed by issuing authorized queries and running the same function on the returned results.

If such metadata is exposed to search parties, $f_S$ must additionally satisfy the \textit{separability} property:
\[
MS_S = \biguplus_{i=1}^u MS_{S,w_i}, \qquad
MS_{S,w_i} = f_S\big((w_i, q_1, \text{Result}(w_i, q_1, S)), \dots \big),
\]
ensuring that each search party $w_i$ can only view the portion of $MS_S$ derivable from their own authorized queries.




\paragraph{\textbf{(B) Overlay-Level Conservativity:}} Overlay network nodes (e.g., indexing routers or federated search controllers) may maintain metadata derived from multiple servers to support query routing or result merging. This system-level metadata $MS_N(t)$ is computed incrementally via a deterministic aggregation function:
\[
MS_N(t) = f_{N,S} \left( t, \{(I_j, MS_{S_j}(t_j))\}_{j=1}^{v} \right),
\quad \text{where } t_j \le t.
\]
where each pair $(I_j, MS_{S_j}(t_j))$ denotes a snapshot of the server-level
metadata from server $S_j$, identified by $I_j$, available at time $t_j \le t$.
We allow $t_1, \dots, t_v$ to be earlier than $t$, reflecting that updates to
system-level metadata may be delayed due to bandwidth or aggregation complexity.


If overlay-level metadata is exposed to search parties, $f_{N,S}$ must satisfy the same \textit{separability} constraint as $f_S$, ensuring that the portion of $MS_N(t)$ visible to a search party $W_i$ can be derived solely from queries that $W_i$ is authorized to issue.

\begin{itemize}
    \item $f_{N,S}$ is an aggregation function, not an oracle. It does not synthesize or infer new information but merges observable metadata from server-local contexts.
    \item Overlay-level metadata is thus constrained by the same conservativity principle: it cannot amplify the visibility of any search party beyond what local server metadata and query results already reveal.
\end{itemize}


\paragraph{\textbf{Explicit Reconstruction Algorithms}}
We now make explicit the reconstruction functions $f_S$ and $f_{N,S}$ referenced above, including their input and output types.

Let $w \in \mathcal{W}$ denote a WebID, $q \in \mathcal{Q}$ a query term, $S$ a server, and $P$ denote a pod URL (pod identifier). Let $\text{Result}(w,q,S) \subseteq \mathcal{P}$ denote the set of pod URLs hosted by $S$ that match $q$ and are authorised for $w$.

A server-local authorised result record is a tuple
\[
\langle w, q, S, R \rangle \quad \text{where } R = \text{Result}(w,q,S).
\]

In what follows, we instantiate the reconstructed metadata statistics using only the cardinality of the authorised result set, i.e., $|R|$, for simplicity. However, the same reconstruction pattern can be extended to compute additional server-level or system-level statistics from authorised results, such as term-frequency summaries, collection-size summaries, or other statistics required for source selection and ranking.

\paragraph{\textbf{Server-level reconstruction function $f_S$.}}
The function $f_S$ maps authorised result records produced by a server $S$ to server-level source-selection metadata partitioned by WebID.

\begin{algorithm}[H]
\caption{$f_S$: Server-Level Metadata Reconstruction}
\label{alg:fS}
\begin{algorithmic}[1]
\Require $\textsf{Log}_S$: authorised result records $\langle w,q,S,R\rangle$
\Ensure $\textsf{MS}_S$: server metadata indexed by WebID and query
\State initialise empty map $\textsf{MS}_S$
\ForAll{$\langle w,q,S,R\rangle \in \textsf{Log}_S$}
    \State $\textsf{MS}_S[w][q].\textsf{Pods} \gets \textsf{MS}_S[w][q].\textsf{Pods} \cup R$
    \State $\textsf{MS}_S[w][q].\textsf{Count} \gets |\textsf{MS}_S[w][q].\textsf{Pods}|$
\EndFor
\State \Return $\textsf{MS}_S$
\end{algorithmic}
\end{algorithm}

We assume that for any pair $(w,q)$ not yet present in $\textsf{MS}_S$,
the corresponding entry is implicitly initialized with
$\textsf{Pods} = \emptyset$ and $\textsf{Count} = 0$.

Thus, for all $w,q$, $\textsf{MS}_S[w][q]$ is exactly the union and cardinality of authorised results $\text{Result}(w,q,S)$.

\paragraph{\textbf{Overlay-level reconstruction function $f_{N,S}$.}}

The overlay aggregates server-level metadata without introducing new information.
The aggregation logic of $f_{N,S}$ depends only on the content of the server-level metadata; time parameters and server identifiers are carried to model network asynchrony and routing, but do not affect correctness.

\begin{algorithm}[H]
\caption{$f_{N,S}$: Overlay-Level Metadata Reconstruction}
\label{alg:fNS}
\begin{algorithmic}[1]
\Require $\{(I_j, t_j, \textsf{MS}_{S_j}(t_j))\}_{j=1}^{v}$ server metadata snapshots
\Ensure $\textsf{MSN}$ overlay metadata indexed by WebID and query
\State initialise empty map $\textsf{MSN}$
\ForAll{$(I_j, t_j, \textsf{MS}_{S_j}(t_j))$}
  \ForAll{$w,q$ such that $\textsf{MS}_S[w][q].\textsf{Count} > 0$}
    \State $\textsf{MSN}[w][q].\textsf{Servers} \gets \textsf{MSN}[w][q].\textsf{Servers} \cup \{I_j\}$
  \EndFor
\EndFor
\ForAll{$w,q$}
  \State $\textsf{MSN}[w][q].\textsf{Count} \gets |\textsf{MSN}[w][q].\textsf{Servers}|$
\EndFor
\State \Return $\textsf{MSN}$
\end{algorithmic}
\end{algorithm}

The overlay metadata therefore records, per WebID, only which servers already
returned authorised non-empty results.

\paragraph{Note.}
In general, the aggregation policy implemented by $f_{N,S}$ may vary across
overlay networks $N$ and across overlay node roles (e.g., routing nodes versus
aggregation nodes).
The function $f_{N,S}$ is defined at the overlay level and is independent of
individual servers, operating only on server-level metadata snapshots that are
already access-controlled.

\paragraph{\textbf{Separability per WebID Example}}
\label{sec:pg4-example}

Consider two servers $S_1,S_2$ and three pods:
$S_1$ hosts $\{P_A,P_B\}$ and $S_2$ hosts $\{P_C\}$.
Let $q$ be a query keyword and consider two WebIDs $w_1$ and $w_2$.

Assume authorised query execution yields:
\[
\text{Result}(w_1,q,S_1)=\{P_A\}, \quad
\text{Result}(w_2,q,S_1)=\{P_B\}
\]
\[
\text{Result}(w_1,q,S_2)=\{P_C\}, \quad
\text{Result}(w_2,q,S_2)=\emptyset.
\]

Applying $f_S$ yields:
\[
\textsf{MS}_{S_1}[w_1][q]=(\{P_A\},1), \quad
\textsf{MS}_{S_1}[w_2][q]=(\{P_B\},1),
\]
\[
\textsf{MS}_{S_2}[w_1][q]=(\{P_C\},1), \quad
\textsf{MS}_{S_2}[w_2][q]=(\emptyset,0).
\]

Applying $f_{N,S}$ yields:
\[
\textsf{MSN}[w_1][q]=(\{S_1,S_2\},2), \quad
\textsf{MSN}[w_2][q]=(\{S_1\},1).
\]

Crucially, metadata for $w_1$ is computable solely from authorised results of $w_1$, and analogously for $w_2$.
No information about pods or servers visible only to $w_2$ is derivable from $\textsf{MSN}[w_1]$, demonstrating separability per WebID.

\medskip
\noindent\textbf{Security Implications of metadata conservativity:}  
This property guarantees that:
\begin{itemize}
    \item No search party can gain indirect knowledge about inaccessible text resources or keywords via metadata structures.
    \item Metadata cannot be used to reverse-engineer hidden data or bypass access control, even if the metadata is inspected directly.
    \item Metadata structures are non-amplifying: they reflect what is already observable through authorized access, not more.
\end{itemize}

\medskip
\noindent\textbf{Example:}  
Suppose WebID $w$ submits a query for the keyword \texttt{cancer} and receives results from pods $P_1$, $P_4$, and $P_7$. The server may record this internally as a metadata tuple (\texttt{cancer}, 3), or update a Bloom filter with those pod references. However, this metadata is entirely reconstructible from the result set that $w$ already received, and would not have been created unless the Search App produced those results through access-respecting evaluation. This ensures that even if adversaries inspect metadata, they \textit{cannot learn anything they could not already infer from aggregating results retrieved from their permitted queries}. Therefore, Metadata is non-amplifying and access-bounded.



It is worth mentioning that the metadata separability and conservativity guarantees in ESPRESSO differ fundamentally from \textit{Differential Privacy} (DP)~\cite{zhao2022survey} as they restrict information flow but do not introduce statistical noise to obscure aggregate properties. As a result, statistical inference attacks (e.g., keyword frequency estimation or correlated inference) remain possible within authorized visibility, whereas DP would explicitly mitigate such inference risks by perturbing metadata or responses.

\subsection{Mapping Adversary Goals to ESPRESSO Privacy Guarantees}

In Table~\ref{tab:goals-vs-guarantees}, we summarize how each privacy guarantee addresses the different adversarial goals introduced in the previous section. It highlights how ESPRESSO’s privacy guarantees jointly constrain different inference channels rather than relying on a single defense. For example, access-bounded guarantees (PG1 and PG2) provide strong protection against direct disclosure and cross-user leakage, while metadata-oriented guarantees (PG3 and PG4) limit amplification of information through source selection and aggregation. Goals marked with $\sim$ reflect residual statistical inference that remains possible within authorized visibility scopes, emphasizing that ESPRESSO limit information exposure but does not eliminate all correlation-based inference.


\begin{table}[t]
\caption{Relationship between Information Disclosure adversary goals and privacy guarantees}
\label{tab:goals-vs-guarantees}

\resizebox{\linewidth}{!}{%
\begin{tabular}{p{4cm} c c c c p{5cm}}
\toprule
Adversary Goal & PG1 & PG2 & PG3 & PG4 & Comments / Protection Mechanism\\
\midrule

(G1) Membership Inference 
& \checkmark &  & \checkmark &  
& Query results are restricted to authorized visibility; Bloom filters prevent reliable negative responses.\\

(G2) Access Pattern Inference 
& \checkmark &  &  & \checkmark 
& Only result-bearing pods within the visibility scope are revealed; metadata is derived solely from visible output.\\

(G3) Keyword Frequency Estimation 
&  & \checkmark & \checkmark &  
& Per-user index scoping hides global keyword frequencies; Bloom filters introduce controlled false positives.\\

(G4) Index Reconstruction 
&  & \checkmark & \checkmark & \checkmark 
& Isolated per-user indices prevent cross-user leakage; metadata reflects only observable results.\\

(G5) Indirect Inference via Correlation 
& \checkmark & \checkmark & \textasciitilde & \textasciitilde 
& Direct keyword access is hidden, but statistical inference through result correlations is not explicitly prevented.\\

(I1) Direct Identification 
& \checkmark & \checkmark &  &  
& Search results include only authorized references; unauthorized identity disclosure is prevented.\\

(I2) Re-identification via Quasi-identifiers 
& \checkmark & \checkmark & \textasciitilde &  
& Index isolation and limited metadata reduce linkage risk; no formal anonymization guarantees are enforced.\\

\bottomrule
\end{tabular}%
}
\end{table}


\section{ESPRESSO Scope and Limitations.} \label{sec:scope-limitations}
The privacy guarantees of ESPRESSO are defined under explicit design and threat-model boundaries that balance privacy protection, system performance, and deployability in real-world Solid environments.

\begin{itemize}
    \item \textbf{No Differential Privacy:} ESPRESSO does not introduce statistical noise into metadata or query results. Incorporating differential privacy mechanisms is orthogonal to our design and left for future work.
        
    \item \textbf{No Cryptographic Query Processing:} Index construction, lookup, and ranking are performed in plain text. ESPRESSO does not rely on secure multiparty computation, homomorphic encryption, or oblivious RAM.
    
    \item \textbf{Hosting Server Trust Boundary:} ESPRESSO assumes \textbf{correctness trust} for hosting servers, i.e., correct enforcement of authentication, access control, and pod-level index isolation. However, the \textbf{Privacy trust} is not assumed within ESPRESSO. Even protocol-compliant servers may be honest-but-curious and attempt to infer information from query execution or metadata, and are therefore included in the adversary model. Fully malicious or non-compliant servers are out of scope for this work. Addressing such servers through verifiable enforcement, trusted execution, or cryptographic query processing is left to future work.

    \item \textbf{Limited Collusion Model:} Our analysis considers independently acting adversaries. Collusion across WebIDs, servers, or infrastructure layers may increase inference power, but does not bypass access control or index isolation. Stronger collusion resistance is left to future work.

    \item \textbf{No Query Obfuscation:} ESPRESSO does not employ dummy queries, traffic padding, or access-pattern obfuscation. Consequently, traffic analysis and timing-based inference are not explicitly mitigated. However, with the privacy guarantees (PG1-PG4), observing results of multiple queries will not reveal any statistical data beyond the access rights of the search party.

    \item \textbf{Authentication and Session Security:} ESPRESSO delegates authentication, session lifecycle management, delegated agent authorization, and revocation handling to the Solid ecosystem. Attacks exploiting session fixation, token replay, misuse of delegated agents, or delayed revocation are considered outside the scope of this work and are assumed to be mitigated by Solid-compliant identity providers and servers. ESPRESSO’s threat model focuses on the privacy implications of decentralized search under the assumption of correct authentication enforcement.
\end{itemize}

These boundaries are intentional and reflect a design choice to provide strong access-bounded privacy guarantees without compromising scalability or compatibility with existing Solid deployments. Extending ESPRESSO with noise-based defenses, cryptographic enforcement, or collusion-resilient mechanisms is a promising direction for future research.

\section{Conclusion \& Future Directions} \label{sec:conclusion-future}
Our threat model reflects practical risks in real-world decentralized environments. The guarantees offered by ESPRESSO, including access and index isolation and the conservative, separable treatment of source-selection metadata, establish a robust foundation for secure decentralized search.
While ESPRESSO currently relies on separability and conservativity for access control, differential privacy (DP) could be introduced in future versions to mitigate statistical inference risks. DP noise could be added at the pod level, server level, or overly network level, trading some accuracy for stronger protection against membership and frequency inference. Other future enhancements may include oblivious query execution and encrypted indexing to provide formal semantic security. A deeper analysis of authentication-layer threats—such as delegated agent misuse, session life cycle management, and revocation propagation-remains an important direction for future work, particularly as Solid deployments increasingly rely on automated agents and long-lived access tokens.


\bibliographystyle{ACM-Reference-Format}
\bibliography{sample}

\end{document}